\documentstyle[aps,twocolumn]{revtex}
\begin{document}
\draft
\title{New Gauge Supergravity in Seven and Eleven Dimensions}
\author{Ricardo Troncoso and Jorge Zanelli}
\address{Centro de Estudios Cient\'{\i}ficos de Santiago,
Casilla 16443, Santiago 9, Chile\\
Departamento de F\'{\i}sica, Universidad de Santiago de Chile, Casilla 307,
Santiago 2, Chile.}
\maketitle

\begin{abstract}
Locally supersymmetric systems in odd dimensions whose Lagrangians are
Chern-Simons forms for supersymmetric extensions of anti-de Sitter gravity are discussed. The construction is illustrated for $D=7$ and 11. In seven dimensions the theory is an $N=2$ supergravity whose fields are the vielbein ($e_{\mu}^a$), the spin connection ($\omega_{\mu}^{ab}$), two gravitini ($\psi_{\mu}^i$) and an $sp(2)$ gauge connection ($a_{\mu j}^i$). These fields form a connection for $osp(2|8)$. In eleven dimensions the theory is an $N=1$ supergravity containing, apart from $e_{\mu}^a$ and $\omega_{\mu}^{ab}$, one gravitino $\psi_{\mu}$, and a totally antisymmetric fifth rank Lorentz tensor one-form, $b_{\mu}^{abcde}$. These fields form a connection for $osp(32|1)$. The actions are by construction invariant under local supersymmetry and the algebra closes off shell without requiring auxiliary fields. The $N=2^{[D/2]}$-theory can be shown to have nonnegative energy around an AdS background, which is a classical solution that saturates the Bogomolnyi bound obtained from the superalgebra.
\end{abstract}

\preprint{\begin{tabular}{l} hep-th/9710180\end{tabular}}

{\bf Introduction.-} In recent years, M-Theory has become the preferred description for the underlying structure of string theories \cite{witten,schwarz}. However, although many features of M-Theory have been identified, still no action principle for it has been given.  

Some of the expected features of M-theory are: (i)Its dynamics should somehow exhibit a superalgebra in which the anticommutator of two supersymmetry generators coincides with the AdS superalgebra in 11 dimensions $osp(32|1)$ \cite{townsend}; (ii)The low energy regime should be described by an eleven dimensional supergravity of new type which should stand on a firm geometric foundation in order to have an off-shell local supersymmetry \cite{NG}; (iii) The perturbation expansion for graviton scattering in M-theory has recently led to conjecture that the new supergravity lagrangian should contain higher powers of curvature \cite{GV}.  Since supersymmetry and geometry are two essential ingredients, most practitioners use the standard 11-dimensional  Cremer-Julia-Scherk ({\bf CJS}) supergravity \cite{CJS} as a good approximation to M-Theory, in spite of the conflict with points (i) and (iii).  In this letter we present a family of supergravity theories which, for D=11, exhibits all of the above features.

In spite of its improved ultraviolet behavior, the renormalizability of standard supergravity beyond the first loops has remained elusive.  It is not completely absurd to speculate that this could be related to the fact that supersymmetry transformations of the dynamical fields form a closed algebra only on shell. An on-shell algebra might seem satisfactory in the sense that the action is nevertheless invariant under local supersymmetry. It is unsatisfactory, however, because it means that the propagating fields neither belong to an irreducible representation of the supergroup nor do they transform as gauge connections. This precludes a fiber-bundle interpretation of the theory, as is the case with standard Yang-Mills gauge theories, where this interpretation is crucial in proving renormalizability. In order to accomodate the supergravity multiplets in tensor representations, it is usually necessary to introduce a host of auxiliary fields . This is a highly nontrivial issue in general and has often remained an unsolvable problem \cite{pvn,rivelles}.

Still, there is a handful of supergravities whose superalgebras close off
shell without requiring auxiliary fields: Anti-de Sitter ({\bf AdS}) in $D=3$ \cite{AT}, and $D=5$ \cite{chamseddine}; Poincar\'{e} in $D=3$ \cite{HIPT}, and in general for $D=2n-1$ \cite{BTrZ}. These are genuine gauge systems for graded Lie algebras and therefore make interesting candidates for renormalizable theories of gravity.

Here, we present a family of supergravity theories in $2n-1$ dimensions whose Lagrangians are Chern-Simons ({\bf CS}) forms related to the $n$-th Chern character of a supergroup in $2n$ dimensions \cite{footnote1}.  As anticipated in the pioneering work of Ref. \cite{CJS}, and underlined by other authors \cite{vV,townsend}, $D=11$ supergravity should be related to a gauge system for the group $OSp(32|1)$, a symmetry which is not reflected in the CJS theory.  The theory proposed here, for $D=11$ turns out to be naturally a gauge system for this group.

{\bf Gauge Gravity.-} Our aim is to construct a locally supersymmetric theory whose generators form a closed off-shell algebra. One way to ensure this by construction is by considering a gauge theory for a graded Lie algebra, that is, one where the structure constants are independent of the fields, and of the field equations.

In order to achieve this, we relax two implicit assumptions usually made about the purely gravitational sector: {\bf (i)} gravitons are described by the Hilbert action, and, {\bf (ii)} torsion does not contain independently propagating degrees of freedom.

The first assumption is historical and dictated by simplicity but in no way
justified by need. In fact, for $D>4$ the most general action for gravity
--generally covariant and with second order field equations for the metric--
is a polynomial of degree $[D/2]$ in the curvature, first discussed by
Lanczos \cite{lanczos} for $D=5$ and, in general, by Lovelock \cite
{lovelock,zumino}. This action contains the same degrees of freedom as the
Hilbert action \cite{T-Z} and is the most general low-energy effective
theory of gravity derived from string theory \cite{zwiebach}.

Assumption {\bf (ii)} is also motivated by simplicity. It means that the
spin connection is not an independent field. Elimination of $\omega _{b}^{a}$ in favor of the remaining fields, however, spoils the possibility of interpreting the local translational invariance as a gauge symmetry of the action. In other words, the spin connection and vielbein --the soldering
between the base manifold and the tangent space-- cannot be identified as
components of the connection for local Lorentz rotations and translations,
respectively, as is the case in $D=3$. Thus, the Einstein-Hilbert theory in $D=\geq 4$ cannot be formulated purely as a gauge theory on a fiber bundle.

For a generic gravitational action in $D>4$, $\frac{\delta S}{\delta \omega}= 0$ cannot be solved for $\omega $ in terms of $e$. This implies
that {\em even classically}, $\omega $ and $e$ should be assumed as
dynamically independent fields and torsion necessarily contains many propagating degrees of freedom \cite{B-G-H}. These degrees of freedom                                                                              are described by the contorsion tensor, $k^{ab}_{\mu} :=\omega^{ab}_{\mu}- \omega^{ab}_{\mu}(e)$. Thus, the restriction to theories with nonpropagating torsion would be a severe truncation in general.

The LL theory, which includes as a particular case the EH system, is by construction invariant under Lorentz rotations in the tangent space. For $D=2n-1$, however, there is a special choice of the coefficients in the LL lagrangian which extends this invariance into an AdS symmetry \cite{chamseddine+}:  $L_{G\;2n-1}^{AdS}= \sum_{p=0}^{n-1}\alpha _{p}L^{p}$, where $\alpha _{p}=\kappa (D-2p)^{-1} (_{\;\;p}^{n-1}) l^{2p-D}$, and
\begin{equation}
L^p_G=\epsilon _{a_{1}\cdots a_{D}}R^{a_{1}a_{2}}\cdots
R^{a_{2p-1}a_{2p}}e^{a_{2p+1}}\cdots e^{a_{D}}.  \label{L-AdS}
\end{equation}
Here wedge product is understood and the subscript ``$G$" denotes a Lagrangian for torsion-free gravity. The constant $l$ has dimensions of length and its purpose is to render the action dimensionless allowing the interpretation of $\omega $ and $e$ as components of the AdS connection \cite{JJG} 
\begin{equation}
W^{AB}=\left[\begin{array}{cc}
\omega ^{ab} & e^{a}/l \\ 
-e^{b}/l & 0
\end{array}
\right], 
\label{W}
\end{equation}
where $A,B=1,...D+1$. The Lagrangian (\ref{L-AdS}) is an AdS-CS form in the
sense that its exterior derivative is the Euler class, 
\begin{equation}
dL^{AdS}_{G\;2n-1}=\kappa \epsilon _{A_{1}\cdots A_{2n}}R^{A_{1}A_{2}}\cdots
R^{A_{2n-1}A_{2n}},  \label{E}
\end{equation}
where $R^{AB}$ is the AdS curvature and $\kappa$ is quantized \cite{z} (in the following we will set $\kappa= l=1$).

Although torsion in general appears in the field equations, it has not been necessary to introduce it explicitly in the action until now. As shown in \cite{mz}, the LL actions can be extended to allow for torsion so that for each dimension there is a unique set of possible additional terms to be considered. Like in the pure LL theories, the most general action contains a large number of arbitrary constants, and again as in the LL case, their number can be reduced to two if the Lorentz invariance is enlarged to AdS symmetry.  

The key point, however, is that torsion terms are necessary in general in order to further extend the AdS symmetry of the action into a supersymmetry for $D>3$. The reason for this is analysed in the discussion.

Let us now briefly examine how torsion appears in an AdS-invariant theory. The idea is best understood in 2+1 dimensions. For $D=3$, apart from the standard action(\ref{E}), there is a second CS form for the AdS group. The exterior derivative of this ``exotic'' lagrangian is the Pontryagin form in 4 dimensions ($2^{nd}$ Chern character for $SO(4)$). This alternative CS form is, \cite{exotic}
\begin{equation}
L^{AdS}_{T\;3}= L^*_3(\omega) +2e_a T^a,
\label{exotic}
\end{equation}
where $dL^*_{2n-1} (\omega^a_b) = Tr[(R^a_b)^n]$ is the $n$-th Chern character (see, e.g. \cite{nakahara}). Similar exotic actions, associated to the Chern characters in $4k$ dimensions, exist in $D=4k-1$. Since the ($2n+1$)-th gravitational Chern characters vanish, there are no exotic actions in $D=4k+1$. For $D=4k-1$, the number of possible exotic forms grows as the partitions of $k$. As we shall see below, we will be interested in one particular combination of these forms, which in the spinorial representation of $SO(4k)$ can be written as \cite{Tr-Z} 
\begin{equation}
dL_{T\;4k-1}^{AdS}=-Tr[(\frac{1}{4}R^{AB}\Gamma _{AB})^{2k}].  
\label{LT}
\end{equation}
It is important to note that in this Lagrangian, as well as in  (\ref{exotic}), torsion appears explicitly. For example, in seven dimensions one finds
\begin{eqnarray*}
&&L_{T\;7}^{AdS}= L_7^*(\omega )-\!\frac{3}{4}(R^a\,_bR^b\,_a + 2[T^aT_a \!-\! R^{ab}e_a e_b]) L^*_3(\omega) \nonumber \\
&&-(\frac{3}{2}R^a\,_bR^b\,_a + 2 T^aT_a -\! 4R^{ab}e_a e_b)T^a e_a + 4T_a R^a\,_bR^b\,_a e^c. \nonumber
\end{eqnarray*}

The CS lagrangian (\ref{LT}) represents a particular choice of coefficients so that the local Lorentz symmetry is enlarged to AdS invariance.  In general, a Chern-Simons D-form is defined so that its exterior derivative is an invariant homogeneous polynomial of degree $n$ in the curvature, that is, a characteristic class \cite{footnote2}. In the examples above, (\ref{E}) is the CS form for the Euler characteristic class $2n$-form, while the exotic lagrangians are related to different combinations of Chern characters.  Thus, a generic CS action in $2n-1$ dimensions for a Lie algebra $g$ can be written as 
\begin{equation}
dL_{2n-1}^{g}=<\mbox{{\bf F}}^{n}>,
\label{F^n}
\end{equation}
where $<$ $>$ stands for a multilinear function in the Lie algebra $g$, invariant under cyclic permutations such as Tr or STr. The problem of finding all possible CS actions for a given group is equivalent to finding all possible invariant tensors of rank $n$ in the algebra. This is in general an open problem, and for the groups relevant for supergravity discussed below (e.g., $OSp(32|1)$) the number of invariant tensors can be rather large.  Most of these invariants, however, give rise to bizarre lagrangians and the real problem is to find the appropriate invariant that describes a sensible theory.

The R.H.S. of (\ref{LT}) is a particular form of (\ref{F^n}) in which $<$ $>$ is the ordinary trace over spinor indices. Other possibilities of the form $<${\bf F}$^{n-p}><${\bf F}$^{p}> $, are not used in our construction as they would not lead to the minimal supersymmetric extensions of AdS containing the Hilbert action. In the supergravity theories discussed below, the gravitational sector is given by $\pm \frac{1}{2^n}L_{G\;2n-1}^{AdS} -\frac{1}{2}L_{T\; 2n-1}^{AdS}$ \cite{footnote3}.

{\bf Gauge Supergravity.-} The supersymmetric extension of a theory invariant under AdS requires new bosonic generators to close the superalgebra \cite{vV}.  In standard supergravities, Lorentz tensors of rank higher than two were usually excluded from the superalgebra on the grounds that elementary particle states of spin higher than 2 are inconsistent \cite{nahm}. However, this does not rule out the relevance of those tensor generators in theories of extended objects \cite{AGIT}.

In \cite{BTrZ}, we discussed family of theories in odd dimensions whose algebra contains the Poincar\'{e} generators . The anticommutator of the supersymmetry generators is a combination of a translation plus a tensorial ``central" extension, 
\begin{equation}
\{Q^{\alpha },\bar{Q}_{\beta }\}=-i(\Gamma ^{a})_{\beta }^{\alpha
}P_{a}-i(\Gamma ^{abcde})_{\beta }^{\alpha }Z_{abcde}.  \label{supertrans}
\end{equation}
This algebra gives rise to supergravity theories with off-shell Poincar\'e superalgebra. The existence of these theories suggests that there should be similar supergravities based on the AdS symmetry. It is our purpose here to present these theories.

{\bf Superalgebra and Connection.-} The smallest superalgebra containing the AdS algebra in the bosonic sector is found following the same approach as in \cite{vV}, but lifting the restriction of $N=1$ \cite{Tr-Z}. The result, for $D>3$ is: \\
\begin{center}
\begin{tabular}{|l|c|c|c|}
\hline
D & S-Algebra & Conjugation Matrix & Internal Metric \\ \hline
$8k-1$ & $osp(N|m)$ & $C^{T}=C$ & $u^{T}=-u$ \\ \hline
$8k+3$ & $osp(m|N)$ & $C^{T}=-C$ & $u^{T}=u$ \\ \hline
$4k+1$ & $su(m|N)$ & $C^{\dag }=C$ & $u^{\dag }=u$ \\ \hline
\end{tabular}
\end{center}
In each of these cases, $m=2^{[D/2]}$ and the connection takes the form 
\begin{eqnarray}
\mbox{{\bf A}}&=& \frac{1}{2}\omega ^{ab}J_{ab}+ e^{a}J_{a} + \frac{1}{r!} b^{[r]}Z_{[r]}+ \nonumber \\
&&\frac{1}{2}(\bar{\psi}^i Q_i -\bar{Q}^i \psi_i) + \frac{1}{2}a_{ij}M^{ij}.  
\label{A}
\end{eqnarray}

The generators $J_{ab},J_{a}$ span the AdS algebra,  $Q_{\alpha}^i$ generate (extended) supersymmetry transformations, and $[r]$ denotes a set of $r$ antisymmetrized Lorentz indices. The $Q's$ transform as vectors under the action of $M_{ij}$ and as spinors under the Lorentz group. Finally, the $Z$'s complete the extension of AdS into the larger algebras $so(m)$, $sp(m)$ or $su(m)$.  

In (\ref{A}) $\bar{\psi}^{i}=\psi _{j}^{T}Cu^{ji}$ ($\bar{\psi}^{i}=\psi _{j}^{\dag}Cu^{ji}$ for $D=4k+1$), where $C$ and $u$ are given in the table above. These algebras admit $(m+N)\times (m+N)$ matrix representations \cite{freund}, where the $J$ and $Z$ have entries in the $m\times m$ block, the $M_{ij}$'s in the $N\times N$ block, while the fermionic generators $Q$ have entries in the complementary off-diagonal blocks.

Under a gauge transformation, {\bf A} transforms by $\delta${\bf A}$= \nabla \lambda $, where $\nabla$ is the covariant derivative for the connection {\bf A}. In particular, under a supersymmetry transformation, $\lambda= \bar{\epsilon}^{i}Q_{i} -\bar{Q}^{i}\epsilon _{i} $, and 
\begin{equation}
\delta _{\epsilon }\mbox{{\bf A}}=\left[ 
\begin{array}{cc}
\epsilon ^{k}\bar{\psi}_{k}-\psi ^{k}\bar{\epsilon}_{k} & D\epsilon _{j} \\ 
-D\bar{\epsilon}^{i} & \bar{\epsilon}^{i}\psi _{j}-\bar{\psi}^{i}\epsilon
_{j} \end{array}
\right] ,  
\label{delA}
\end{equation}
where $D$ is the covariant derivative on the bosonic connection, $D\epsilon _{j}=(d+\frac{1}{2} [e^a\Gamma_a + \frac{1}{2}\omega^{ab}\Gamma_{ab} + \frac{1}{r!}b^{[r]}\Gamma _{[r]}])\epsilon_{j} -a_{j}^{i}\epsilon _{i}$.

{\bf D=7} The smallest AdS superalgebra in seven dimensions is $osp(2|8)$. The connection (\ref{A}) is {\bf A} =$\frac{1}{2}\omega ^{ab}J_{ab}+e^{a}J_{a}+ \bar{Q}^{i}\psi _{i}+\frac{1}{2}a_{ij}M^{ij}$, where $M^{ij}$ are the generators of $sp(2)$. In the representation given above, the bracket $<$ $>$ is the supertrace and, in terms of the component fields appearing in the connection, the CS form is
\begin{eqnarray}
L_7^{osp(2|8)}(\mbox{{\bf A}})&=& 2^{-4}L_{G\;7}^{AdS}(\omega,e)-\frac{1}{2} L_{T\; 7}^{AdS}(\omega,e) \nonumber \\
&& -L_7^{*Sp(2)}(a)+L_{F}(\psi ,\omega ,e,a).
\end{eqnarray}
Here the fermionic Lagrangian is 
\begin{eqnarray*}
L_F &=&4\bar{\psi}^j(R^{2}\delta_j^i + Rf_j^i +(f^2)_j^i)D\psi_i \nonumber \\ & &+4(\bar{\psi}^i \psi_j)[(\bar{\psi}^j\psi_k)(\bar{\psi}^k D\psi_i) -\bar{\psi}^j(R\delta_i^k + f_i^k)D\psi_k] \\
&&-2(\bar{\psi}^{i}D\psi _{j})[\bar{\psi}^{j}(R\delta
_{i}^{k}+f_{i}^{k})\psi _{k}+D\bar{\psi}^{j}D\psi _{i}],\nonumber
\end{eqnarray*}
where $f_j^i= da_j^i+a_k^i a_j^k$, and $R=\frac{1}{4}(R^{ab} +e^a e^b) \Gamma_{ab} + \frac{1}{2} T^a \Gamma_a$ are the $sp(2)$ and $so(8)$ curvatures, respectively. The supersymmetry transformations (\ref{delA}) read \\
\begin{tabular}{llll}
\hspace{1cm}& $\delta e^a =\frac{1}{2}\bar{\epsilon}^i\Gamma^a \psi_i $ & \hspace{1cm} & $\delta \omega ^{ab}=-\frac{1}{2}\bar{\epsilon}^i\Gamma^{ab} \psi_i$ \\ 
\hspace{1cm} & $\delta \psi_i =D\epsilon_i$ & \hspace{1cm} & $\delta
a_j^i =\bar{\epsilon}^i \psi_j-\bar{\psi}^i \epsilon_j.$
\label{susy7}
\end{tabular}

{\bf D=11} In this case, the smallest AdS superalgebra is $osp(32|1)$ and the connection is {\bf A} =$ \frac{1}{2}\omega ^{ab}J_{ab} + e^{a}J_{a} + \frac{1}{5!}b^{abcde} J_{abcde} + \bar{Q}\psi$, where $b$ is a totally antisymmetric fifth-rank Lorentz tensor one-form. Now, in terms of the elementary bosonic and fermionic fields, the CS form in (\ref{F^n}) reads 
\begin{equation}
L_{11}^{osp(32|1)}(${\bf A}$) = L_{11}^{sp(32)}(\Omega )+L_{F}(\Omega ,\psi ),  
\label{L11}
\end{equation}
where $\Omega\equiv \frac{1}{2}(e^{a}\Gamma_{a} + \frac{1}{2} \omega^{ab}\Gamma_{ab} + \frac{1}{5!}b^{abcde} \Gamma_{abcde})$ is an $sp(32)$ connection. The bosonic part of (\ref{L11}) can be written as 
\begin{eqnarray}
L_{11}^{sp(32)}(\Omega )&=&2^{-6} L_{G\;11}^{AdS}(\omega ,e) -\frac{1}{2} L_{T\;11}^{AdS}(\omega ,e) + L_{11}^{b}(b,\omega ,e).  
\nonumber
\end{eqnarray}
The fermionic Lagrangian is 
\begin{eqnarray*}
L_{F} &=&6(\bar{\psi}R^{4}D\psi )-3\left[ (D\bar{\psi}D\psi )+(\bar{\psi}
R\psi)\right] (\bar{\psi}R^{2}D\psi ) \nonumber \\
& &-3\left[ (\bar{\psi}R^{3}\psi )+(D\bar{\psi}R^{2}D\psi )\right] (\bar{\psi} D\psi )+ \\
& &2\left[ (D\bar{\psi}D\psi )^{2}+(\bar{\psi}R\psi )^{2}+(\bar{\psi}R\psi) (D\bar{\psi}D\psi )\right] (\bar{\psi}D\psi),
\end{eqnarray*}
where $R=d\Omega +\Omega ^{2}$ is the $sp(32)$ curvature. The supersymmetry transformations (\ref{delA}) read\\
\begin{tabular}{llll}
\hspace{1cm} & $\delta e^{a}=\frac{1}{8}\bar{\epsilon}\Gamma ^{a}\psi $ & 
\hspace{1cm} & $\delta \omega ^{ab}=-\frac{1}{8}\bar{\epsilon}\Gamma ^{ab}\psi $ \\ 
\hspace{1cm} & $\delta \psi =D\epsilon $ & \hspace{1cm} & $\delta b^{abcde}= \frac{1}{8}\bar{\epsilon}\Gamma ^{abcde}\psi.$
\label{susy11}
\end{tabular}

{\bf Discussion.-} The supergravities presented here have two distinctive features: The fundamental field is always the connection {\bf A} and, in their simplest form, these are pure CS systems (matter couplings are discussed below).  As a result, these theories possess a larger gravitational sector, including propagating spin connection. Contrary to what one could expect, the geometrical interpretation is quite clear, the field structure is simple and, in contrast to the standard cases, the supersymmetry transformations close off shell without auxiliary fields.

{\bf A. Torsion.} It can be observed that the torsion lagrangians ($L_T$)are odd while the torsion-free terms ($L_G$) are even under spacetime reflections. The minimal supersymmetric extension of the AdS group in $4k-1$ dimensions requires using chiral spinors of $SO(4k)$ \cite{Gunaydin}. This in turn implies that the gravitational action has no definite parity, but requires the combination of $L_T$ and $L_G$ as described above. In $D=4k+1$ this issue doesn't arise due to the vanishing of the torsion invariants, allowing constructing a supergravity theory based on $L_G$ only, as in \cite{chamseddine}. If one tries to exclude torsion terms in $4k-1$ dimensions, one is forced to allow both chiralities for $SO(4k)$ duplicating the field content, and the resulting theory has two copies of the same system \cite{horava}.

{\bf B. Field content and extensions with N$>$1.}The field content compares with that of the standard supergravities in $D=7$, 11 as follows:\\
\begin{center}
\begin{tabular}{c|c|l|l|}
\cline{2-4}
\hspace{.7cm} & D  & Standard supergravity  & New supergravity \\ \cline{2-4}
 & 7  & $e^a_{\mu}$ $A_{[3]}$ $\psi^{\alpha i}_{\mu}$ $a^i_{\mu j}$ $\lambda^{\alpha}$ $\phi$ & $ e_{\mu }^a$ $\omega ^{ab}_{\mu}$ $\psi_{\mu}^{\alpha i }$ $a_{\mu j}^i$	\\ \cline{2-4}
 & 11 & $e^a_{\mu}$ $A_{[3]}$ $\psi^{\alpha}_{\mu}$ & $e_{\mu}^{a}$ $\omega ^{ab}_{\mu }$ $\psi_{\mu }^{\alpha }$  $b^{abcde}_{\mu }$  \\ 
\cline{2-4}
\end{tabular}
\end{center}

Standard seven-dimensional supergravity is an $N=2$ theory (its maximal extension is N=4), whose gravitational sector is given by Einstein-Hilbert gravity with cosmological constant and with a background invariant under $OSp(2|8)$ \cite{D=7,Salam-Sezgin}. Standard eleven-dimensional supergravity \cite{CJS} is an N=1 supersymmetric extension of Einstein-Hilbert gravity that cannot accomodate a cosmological constant \cite{B-D-H-S}. An $N>1$ extension of this theory is not known.

In the case presented here, the extensions to larger $N$ are straighforward in any dimension. In $D=7$, the index $i$ is allowed to run from $2$ to $2s$, and the Lagrangian is a CS form for $osp(2s|8)$. In $D=11$, one must include an internal $so(N)$ field and the Lagrangian is an $osp(32|N)$ CS form \cite{Tr-Z}. The cosmological constant is necessarily nonzero in all cases.

{\bf C. Spectrum.} The stability and positivity of the energy for the solutions of these theories is a highly nontrivial problem. As shown in Ref. \cite{B-G-H}, the number of degrees of freedom of bosonic CS systems for $D\geq 5$ is not constant throughout phase space and different regions can have radically different dynamical content. However, in a region where the rank of the symplectic form is maximal the theory behaves as a normal gauge system, and this condition is stable under perturbations.  As it is shown in \cite{CTZ}, there exists a nontrivial extension of the AdS superalgebra with one abelian generator for which anti-de Sitter space without matter fields is a background of maximal rank, and the gauge superalgebra is realized in the Dirac brackets. For example, for $D=11$ and $N=32$, the only nonvanishing anticommutator reads
\begin{eqnarray*}
\{Q^i_{\alpha},\bar{Q}^j_{\beta} \} &=& \frac{1}{8}\delta^{ij}\left[ C\Gamma^{a} J_a + C\Gamma^{ab}J_{ab} + C\Gamma^{abcde}Z_{abcde} \right]_{\alpha \beta}\\
& & -M^{ij}C_{\alpha \beta},
\label{Malg}
\end{eqnarray*}
where $M^{ij}$ are the generators of $SO(32)$ internal group.  On this background the $D=11$ theory has $2^{12}$ fermionic and $2^{12}-1$ bosonic degrees of freedom. The (super)charges obey the same algebra with a central extension. This fact ensures a lower bound for the mass as a function of the other bosonic charges \cite{G-H}.

{\bf D. Classical solutions.} The field equations for these theories in terms of the Lorentz components ($\omega$, $e$, $b$, $a$, $\psi$) are spread-out expressions for $<${\bf F}$^{n-1}G_{(a)}> =0$, where $G_{(a)}$ are the generators of the superalgebra. It is rather easy to verify that in all these theories the anti-de Sitter space is a classical solution , and that for $\psi=b=a=0$ there exist spherically symmetric, asymptotically AdS standard \cite{JJG}, as well as topological \cite{pmb} black holes. In the extreme case these black holes can be shown to be BPS states.

{\bf E. Matter couplings.} It is possible to introduce a minimal couplings to matter of the form {\bf A}$\cdot ${\bf J}. For $D=11$, the matter content is that of a theory with (super-) 0, 2, and 5--branes, whose respective worldhistories couple to the spin connection and the $b$ fields.

{\bf F. Standard SUGRA.} Some sector of these theories might be related to the standard supergravities if one identifies the totally antisymmetric part of the contorsion tensor in a coordinate basis, $k_{\mu \nu \lambda}$, with the abelian 3-form, $A_{[3]}$. In 11 dimensions one could also identify the antisymmetric part of $b$ with an abelian 6-form $A_{[6]}$, whose exterior derivative, $dA_{[6]}$, is the dual of $F_{[4]}=dA_{[3]}$. Hence, in $D=11$ the CS theory  possibly contains the standard supergravity as well as some kind of dual version of it.\\

The authors are grateful to S. Deser, G. Gibbons, M. Henneaux, C. Teitelboim and the CECS research group for many enlightening discussions, and to E. Witten for helpful comments. This work was supported in part by grants 1960229, 1970151, 1980788 and 3960009 from FONDECYT (Chile), and 27-953/ZI-DICYT (USACH). Institutional support to CECS from Fuerza A\'erea de Chile and a group of Chilean private companies (Business Design Associates, CGE, CODELCO, COPEC, Empresas CMPC, Minera Collahuasi, Minera Escondida, NOVAGAS and XEROX-Chile) is also acknowledged. J. Z. thanks the John Simon Guggenheim Foundation for support.


\begin{references}
\bibitem{witten} E. Witten, Nucl. Phys. {\bf B443} (1995) 85.
\bibitem{schwarz} J. Schwarz, Phys. Lett. {\bf B367} (1996) 971.
\bibitem{townsend} P. K. Townsend, {\em p-Brane Democracy}, hep-th/9507048
\bibitem{NG} H. Nishino and S. J. Gates, Phys. Lett. {\bf B388} (1996) 504.
\bibitem{GV} M.Green and P. Vanhove, Phys. Lett. {\bf B408}, 122(1997).
\bibitem{CJS} E. Cremmer, B. Julia and J. Scherk, Phys. Lett. {\bf 76B}
(1978) 409.
\bibitem{pvn} P. van Nieuwenhuizen, Phys. Rep. {\bf 68}, 4 (1981).
\bibitem{rivelles} V. O. Rivelles and J. G. Taylor, Phys. Lett. {\bf 104B} (1981) 131; {\bf 121} (1983) 38.
\bibitem{AT} A. Ach\'{u}carro and P. K. Townsend, Phys. Lett. {\bf B180}
(1986) 89.
\bibitem{chamseddine} A.Chamseddine, Nucl.Phys. {\bf B346} (1990) 213.
\bibitem{HIPT} P. S. Howe, J. M. Izquierdo, G. Papadopoulos and P. K.
Townsend, Nucl. Phys. {\bf B467} (1996) 183.
\bibitem{BTrZ} M. Ba\~{n}ados, R. Troncoso and J. Zanelli, Phys. Rev. 
{\bf D54}, (1996) 2605.
\bibitem{footnote1} For simplicity, we will not distinguish between different signatures. Thus, the AdS algebra in $D$ dimensions will be denoted as $so(D+1)$. 
\bibitem{vV} J. W. van Holten and A. Van Proeyen, J. Phys. {\bf A 15}
(1982) 3763.
\bibitem{lanczos} C. Lanczos, Ann. Math. {\bf 39} (1938) 842.
\bibitem{lovelock} D. Lovelock, J. Math. Phys. {\bf 12} (1971) 498.
\bibitem{zumino} B. Zumino, Phys. Rep. {\bf 137} (1986) 109.
\bibitem{T-Z} C. Teitelboim and J. Zanelli, Class. and Quantum Grav. {\bf 4}(1987) L125.
\bibitem{zwiebach} B. Zwiebach, Phys. Lett. {\bf 156B} (1985) 315.
\bibitem{B-G-H} M. Ba\~{n}ados, L. J. Garay and M. Henneaux, Phys. Rev. 
{\bf D53} R593 (1996); Nucl. Phys. {\bf B476} 611 (1996).
\bibitem{chamseddine+}A. Chamseddine, Phys. Lett. {\bf B233} (1989) 113.
\bibitem{JJG} M. Ba\~{n}ados, C. Teitelboim and J. Zanelli, in {\em J. J.
Giambiagi Festschrift}, H. Falomir {\em et al.}(eds.),(World Scientific, Singapore 1991); Phys. Rev. {\bf D 49} (1994) 975.
\bibitem{z} J. Zanelli, Phys. Rev. {\bf D51} (1995) 490.
\bibitem{mz} A. Mardones and J. Zanelli, Class. Quantum Grav. {\bf 8}
(1991) 1545.  
\bibitem{exotic} E. Witten, Nucl. Phys. {\bf B311} (1988) 46.
\bibitem{nakahara} M. Nakahara,{\em Geometry, Topology and Physics} (Adam Hilger, New York 1990).
\bibitem{Tr-Z} R. Troncoso, Tesis Doctoral, Universidad de Chile (1996); and R. Troncoso and J. Zanelli, in the proceedings of the meeting on {\em Quantum Gravity in the Siuthern Cone}, Bariloche, Argentina, January 1998 (to appear).
\bibitem{footnote2} An explicit expression for the CS form in (\ref{F^n}) is $L^g_{2n-1}=\int_{0}^{1}d\lambda<\mbox{{\bf A}}(\lambda d\mbox{{\bf A}} +\lambda^2\mbox{{\bf A}}^2)^{n-1}>$. If the manifold $M$ has no boundary, the action can be written as a topological theory in a $2n$-dimensional space $\Omega$, with $\partial \Omega=M$.
\bibitem{footnote3} The $\pm$ sign correspond to the two choices of inequivalent representations of $\Gamma$'s, which in turn reflect the two chiral representations in $D+1$. As mentioned earlier, the supersymmetric extensions of $L_G$ or $L_T$, require using both chiralities, thus doubling the algebras. Here we choose the + sign, which gives the minimal extension.
\bibitem{nahm} W. Nahm, Nucl. Phys. {\bf B135} (1978) 149. J. Strathdee, Int. J. Mod. Phys. {\bf A 2} (1987) 273.
\bibitem{AGIT} J. A. de Azc\'{a}rraga, J. P. Gauntlet, J. M. Izquierdo and
P. K. Townsend, Phys. Rev. Lett. {\bf 63} (1989) 2443.
\bibitem{freund} P.Freund, {\em Introduction to Supersymmetry} Cambridge
University Press (Cambridge, U.K., 1989).
\bibitem{Gunaydin} M. Gunaydin, hep-th/9803133.
\bibitem{horava} P. Horava, hep-th/9712130.
\bibitem{D=7} P. K. Townsend and P. van Nieuwenhuizen, Phys. Lett. {\bf 125B} (1983) 41.
\bibitem{Salam-Sezgin} A. Salam and E. Sezgin, Phys. Lett. {\bf 126B} (1983) 295. 
\bibitem{B-D-H-S} K. Bautier, S. Deser, M. Henneaux and D.Seminara, Phys. Lett. {\bf B406},(1997) 49.
\bibitem{CTZ} O. Chand\'{\i}a, R. Troncoso and J. Zanelli, "Dynamical Structure of Chern-Simons Supergravities", (in preparation).
\bibitem{G-H} G.W. Gibbons and C.M. Hull, Phys. Lett. {\bf 109B} (1982) 190.
\bibitem{pmb} S. Aminneborg, I. Bengtsson, S. Holst and P. Peldan, Class. Quantum Grav.{\bf 13} (1996) 2707. M. Ba\~{n}ados,Phys. Rev. {\bf D57} (1998), 1068. R.B. Mann, gr-qc/9709039.

\end{references}
\end{document}